\documentclass[aps,prl,showpacs,
superscriptaddress,nofootinbib,floatfix]{revtex4-1}
\usepackage{amsfonts,amssymb,stmaryrd,latexsym,amsmath}
\usepackage{graphicx,subfigure}
\usepackage{times}
\usepackage{slashed}

\begin{document}

\title{\textit{Ab initio} alpha-alpha scattering}

\author{Serdar Elhatisari}
\affiliation{Helmholtz-Institut f\"ur Strahlen- und
             Kernphysik and Bethe Center for Theoretical Physics, \\
             Universit\"at Bonn,  D-53115 Bonn, Germany}
\author{Dean~Lee}
\affiliation{Department~of~Physics, North~Carolina~State~University, Raleigh,
NC~27695, USA}

\author{Gautam~Rupak}
\affiliation{Department~of~Physics~and~Astronomy and HPC$^2$ Center for Computational Sciences, Mississippi~State~University,
Mississippi State, MS~39762, USA}

\author{Evgeny Epelbaum}
\affiliation{Institut~f\"{u}r~Theoretische~Physik~II,~Ruhr-Universit\"{a}t~Bochum,
D-44870~Bochum,~Germany}

\author{Hermann~Krebs}
\affiliation{Institut~f\"{u}r~Theoretische~Physik~II,~Ruhr-Universit\"{a}t~Bochum,
D-44870~Bochum,~Germany}

\author{Timo~A.~L\"{a}hde}
\affiliation{Institute~for~Advanced~Simulation, Institut~f\"{u}r~Kernphysik,
and
J\"{u}lich~Center~for~Hadron~Physics,~Forschungszentrum~J\"{u}lich,
D-52425~J\"{u}lich, Germany}

\author{Thomas~Luu}
\affiliation{Institute~for~Advanced~Simulation, Institut~f\"{u}r~Kernphysik,
and
J\"{u}lich~Center~for~Hadron~Physics,~Forschungszentrum~J\"{u}lich,
D-52425~J\"{u}lich, Germany}
\affiliation{Helmholtz-Institut f\"ur Strahlen- und
             Kernphysik and Bethe Center for Theoretical Physics, \\
             Universit\"at Bonn,  D-53115 Bonn, Germany}

\author{Ulf-G.~Mei{\ss }ner}
\affiliation{Helmholtz-Institut f\"ur Strahlen- und
             Kernphysik and Bethe Center for Theoretical Physics, \\
             Universit\"at Bonn,  D-53115 Bonn, Germany}
\affiliation{Institute~for~Advanced~Simulation, Institut~f\"{u}r~Kernphysik,
and J\"{u}lich~Center~for~Hadron~Physics,~Forschungszentrum~J\"{u}lich,
D-52425~J\"{u}lich, Germany}
\affiliation{JARA~-~High~Performance~Computing, Forschungszentrum~J\"{u}lich,
D-52425 J\"{u}lich,~Germany}

\date{\today}

\begin{abstract}
Processes involving alpha particles and alpha-like nuclei comprise a major part of stellar nucleosynthesis and hypothesized mechanisms for thermonuclear supernovae. In an effort towards understanding alpha processes from first principles, we describe in this letter the first \textit{ab initio} calculation of alpha-alpha scattering.  We use lattice effective field theory to describe the low-energy interactions of nucleons and apply a technique called the adiabatic projection method to reduce the eight-body system to an effective two-cluster system.  We find good agreement between lattice results and experimental phase shifts for $S$-wave and $D$-wave scattering.  The computational scaling with particle number suggests that alpha processes involving heavier nuclei are also within reach in the near future.\end{abstract}


\maketitle

\section{Introduction}

In recent years there has been much progress in {\it{ab initio}} nuclear scattering and reactions involving light nuclei 
\cite{Nollett:2006su,Quaglioni:2008sm,Navratil:2011zs,Navratil:2011sa,Hupin:2013wsa} as well as medium-mass nuclei \cite{Hagen:2012rq,Orlandini:2013eya}.  However, the computational scaling for most numerical methods increases significantly when
the
projectile nucleus has more than a few nucleons. Therefore it remains a challenge to address important alpha processes relevant for stellar astrophysics
such as $^4\rm{He} + {^4\rm{He}}$ scattering and  $^{12}\rm{C} + {^4\rm{He}}$
scattering and radiative capture, as well as  carbon and oxygen burning  in massive star evolution and in thermonuclear
supernovae \cite{Woosley:1973a,Fink:2010a,Shen:2013hpa}.

In this work we describe lattice calculations for which the computational scaling of the $A_1$-body + $A_2$-body problem is roughly $(A_1+A_2)^2$, mild enough to make first principles calculations of alpha processes possible.  We use the formalism of lattice effective field theory 
 \cite{Epelbaum:2009zs,Epelbaum:2011md,Epelbaum:2012qn,Lahde:2015ona} and make use of a technique for elastic scattering and inelastic reactions on the lattice called the adiabatic projection method 
\cite{Rupak:2013aue,Pine:2013zja,Elhatisari:2014lka,Rupak:2014xza,Rokash:2015hra}. In the following we present the first {\it{ab
initio}} calculation of $^4\rm{He} + {^4\rm{He}}$ scattering, going up to next-to-next-to-leading order in chiral effective field theory.  See Ref.~\cite{Epelbaum:2008ga}
for a
recent review of chiral effective field theory. We find good agreement with experimental data for the $S$-wave and $D$-wave phase shifts 
\cite{Heydenburg:1956,Nilson:1958,Tombrello:1963,Afzal:1969}.

\section{Adiabatic projection method}
The adiabatic projection method addresses the cluster-cluster scattering
problem on the lattice by using Euclidean time projection  to
construct an 
effective two-cluster Hamiltonian for the participating clusters.   By Euclidean time projection, we mean multiplication by $\exp(-H\tau)$, where $H$ is the underlying microscopic Hamiltonian. For simplicity of explanation we refer to a continuous Euclidean time
parameter $\tau$, even though the actual lattice calculations use discrete time steps.

\begin{figure}[htb]
\centering
\includegraphics[scale=0.5]{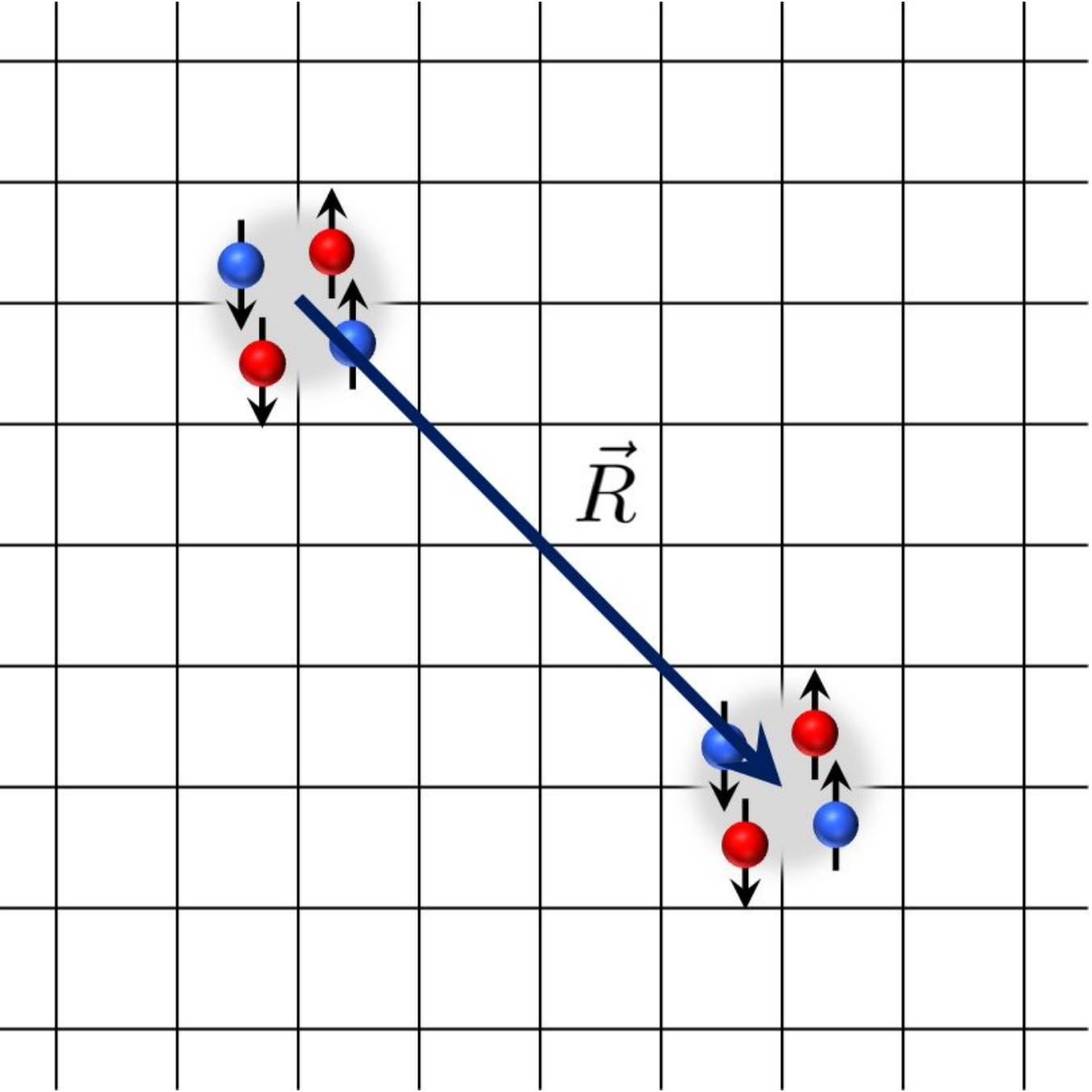}
\caption{Initial state $\vert \vec R \rangle$ of two alpha particles separated by the displacement vector $\vec R$.}
\label{fig:clusters}
\end{figure}

Our starting point is an $L^3$ periodic 
spatial lattice. We take a set of initial two-alpha states $|\vec{R}\rangle$ labeled
by their separation vector $\vec{R}$, as illustrated in
Fig.~\ref{fig:clusters}. We take the initial alpha wave functions to be Gaussian wave packets so that at large separations
they factorize as a tensor product of two individual alpha clusters,
\begin{equation}
|\vec{R}\rangle=\sum_{\vec{r}} |\vec{r}+\vec{R}\rangle_1\otimes|\vec{r}\rangle_2.
\label{eqn:single_clusters}\\ 
\end{equation}
   The summation over $\vec{r}$ produces two-alpha states with total momentum equal to zero. Rather than dealing with a large array of three-dimensional vectors $\vec{R}$,
we find it convenient to project onto spherical harmonics
$Y_{L,L_z}$ with angular momentum quantum numbers $L,L_z,$
\begin{equation}
|R\rangle^{L,L_z} = \sum_{\vec{R'}}Y_{L,L_z}(\hat{R}')\delta_{R,|\vec{R}'|}|\vec{R}'\rangle.
\end{equation}

We then use Euclidean time projection to form dressed cluster
states,
\begin{equation}
\vert R\rangle^{L,L_z}_\tau   =\exp(-H\tau)|R\rangle^{L,L_z}.
\end{equation}  
We are using units where $\hbar$ and $c$ are set to 1. The evolution in Euclidean time automatically incorporates the induced deformation
and
polarization
of the alpha clusters as they come near each other. 

With these dressed cluster states, we compute matrix elements of the full microscopic Hamiltonian
with respect to the dressed cluster states,
\begin{equation}
\left[H_{\tau}\right]^{L,L_z}_{R,R'} =\ ^{L,L_z}_{\quad\;\tau}\langle R\vert H
\vert R'\rangle^{L,L_z}_{\tau}.
\end{equation}
 Since the dressed cluster states are not orthogonal, we construct a norm matrix, 
\begin{equation}
\left[N_{\tau}\right]^{L,L_z}_{R,R'} =\ ^{L,L_z}_{\quad\;\tau}\langle R\vert R'\rangle^{L,L_z}_{\tau}. \label{eqn:norm}
\end{equation}
The radial adiabatic Hamiltonian is defined as the matrix product,
\begin{equation}
\left[ {H^a_{\tau}} \right]^{L,L_z}_{R,R'} = 
\left[N_{\tau}^{-\frac{1}{2}}H_{\tau}
N_{\tau}^{-\frac{1}{2}} \right]^{L,L_z}_{R,R'}.
\label{eqn:Adiabatic-Hamiltonian}
\end{equation}
In the limit of large projection time $\tau$, the spectrum of the adiabatic Hamiltonian reproduces the low-energy finite volume spectrum of the microscropic
Hamiltonian $H$.  In Ref.~\cite{Rokash:2015hra} it was shown that in the asymptotic region where the alpha clusters are widely separated, the adiabatic Hamiltonian reduces to a simple two-cluster Hamiltonian with only infinite-range interactions such as the Coulomb interaction between the otherwise non-interacting clusters.

\section{Lattice calculations and results}

We study $^4\rm{He} + {^4\rm{He}}$ scattering using the same lattice action used to study the Hoyle state of $^{12}$C \cite{Epelbaum:2012qn}, the structure of $^{16}$O \cite{Epelbaum:2013paa}, and other nuclear systems \cite{Lahde:2013uqa}. The spatial lattice spacing is $a = 1.97~{\rm{fm}}$ and the temporal lattice spacing is  $a_t = 1.32~{\rm{fm}}$.    Revisiting these calculations again in the future with different lattice spacings and going to higher order will provide a useful measure of systematic errors in lattice calculations of higher-body nuclear systems \cite{Klein:2015vna}. We should mention that the dependence on lattice spacing has also been explored using model interactions of point-like alpha particles on the lattice \cite{Lu:2014xfa,Lu:2015gfa}.  

We perform projection Monte Carlo simulations with auxiliary fields to compute the matrices 
$\left[ {H_{\tau}} \right]^{L,L_z}_{R,R'}$ and $\left[ {N_{\tau}} \right]^{L,L_z}_{R,R'}$  on a periodic cubic lattice with length $L=16$~fm. See Ref.~\cite{Lee:2008fa} for an overview of methods used in lattice effective field theory. We determine $\left[ {N_{\tau}} \right]^{L,L_z}_{R,R'}$ from calculations
with $L_t$ time steps and $\left[ {H_{\tau}} \right]^{L,L_z}_{R,R'}$ from
calculations with $L_t+1$ time steps.
The projection time $\tau$ equals
the product of $L_t$ and the temporal lattice spacing $a_t$. For these calculations, a new algorithm is used to allow for Monte Carlo updates of the auxiliary fields as well as updates of the alpha cluster positions.

We compute the radial adiabatic Hamiltonian using Eq.~(\ref{eqn:Adiabatic-Hamiltonian}) and extend the radial adiabatic Hamiltonian to a much larger volume with length $120$~fm.  This is done by computing matrix elements of $\left[ {H^a_{\tau}} \right]^{L,L_z}_{R,R'}$ at large separation from single-alpha lattice simulations and then including the   Coulomb interaction between the otherwise non-interacting clusters.  This process also allows us to define a ``trivial'' two-cluster Hamiltonian in which the two alpha clusters are non-interacting except for the infinite-range Coulomb interaction.  

\begin{figure}[htb]
\centering
\includegraphics[scale=1.3]{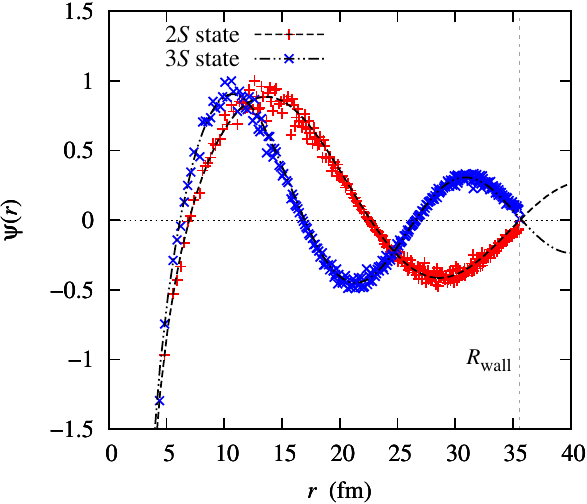}
\caption{$S$-wave scattering wave functions at next-to-next-to-leading
order versus radial distance.}
\label{fig:wave_function}
\end{figure}

With the radial adiabatic Hamiltonian defined in the large $(120~{\rm{fm}})^3$ box, we extract the scattering phase shifts by imposing a hard spherical wall boundary at some chosen radius $R_{\rm{wall}}$ and solving for the standing wave modes. In Fig.~\ref{fig:wave_function} we show $S$-wave radial functions for
two different radial excitations (2$S$ and 3$S$) at next-to-next-to-leading order in chiral effective field theory. We can extract the phase shift
directly by fitting to the asymptotic behavior of the radial wave function
as done in Ref.~\cite{Rokash:2015hra}.  However, we find it more efficient
and accurate
to extract the phase shifts from the energy differences between the spectrum
of the adiabatic Hamiltonian and the trivial two-cluster Hamiltonian. This
technique is discussed in detail in Ref.~\cite{Borasoy:2007vy}.

\begin{figure}[htb]
\centering
\includegraphics[scale=1.0]{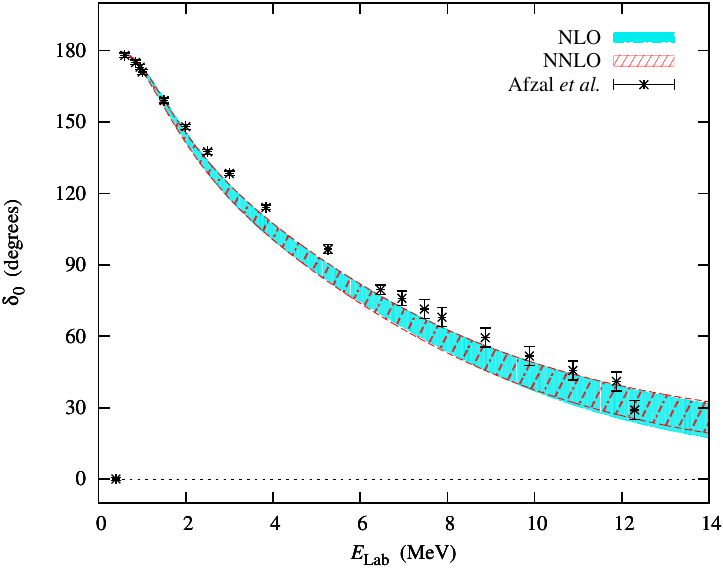}
\caption{$S$-wave phase shifts at NLO and NNLO and comparison with experimental
data \cite{Heydenburg:1956,Nilson:1958,Tombrello:1963,Afzal:1969}.}
\label{fig:S-wave}
\end{figure}

Fig.~\ref{fig:S-wave} shows the phase shifts for $S$-wave scattering versus
laboratory energy at next-to-leading order (NLO) in chiral effective field theory, next-to-next-to-leading order (NNLO), and the comparison with experimental
data \cite{Heydenburg:1956,Nilson:1958,Tombrello:1963,Afzal:1969}.
The results
at leading order (LO) do not include Coulomb effects and so have significantly different
behavior near threshold and are not shown.  We note that the NLO and NNLO phase shifts are
quite similar, and both are in good agreement with experiment. The error bands are estimated from the spread in phase shifts
versus the number of time steps in the projection Monte Carlo calculation, $L_t$. 

\begin{figure}[htb]
\centering
\includegraphics[scale=0.75]{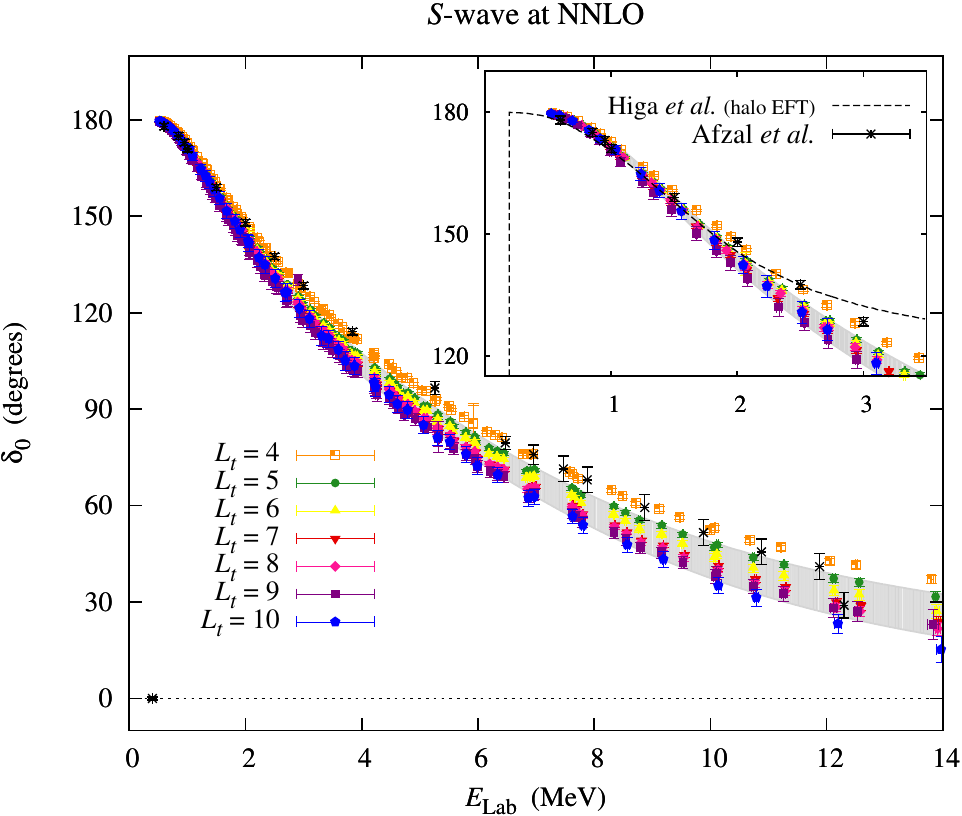}
\caption{$S$-wave phase shifts at NNLO for different numbers of projection
time steps $L_t$ and comparison with experimental
data \cite{Heydenburg:1956,Nilson:1958,Tombrello:1963,Afzal:1969}.  In the inset we show next-to-leading-order calculations using halo effective
field theory \cite{Higa:2008dn}.  }
\label{fig:S-wave_NNLO}
\end{figure}

In Fig.~\ref{fig:S-wave_NNLO} we plot the $S$-wave
phase shifts at NNLO for $L_t=4$ to $L_t =10$.  The error bars are computed using a jackknife analysis of the stochastic errors of the Monte Carlo data.  We find that the calculated $^8$Be ground state is bound at NNLO, though only a small fraction of an MeV away from threshold.  For further comparison, we show in the inset 
next-to-leading-order results using halo effective field theory with point-like alpha particles \cite{Higa:2008dn}. We note that the norm matrix in Eq.~(\ref{eqn:norm}) requires an even number of time steps in order to construct the inner product
of dressed cluster states and so is not strictly defined for odd $L_t$. This explains some systematic differences between the even $L_t$ and odd $L_t$ results.  However, they do appear to agree for large $L_t$ as expected.  The data is not sufficiently precise to allow for a multi-parameter extrapolation to the limit $L_t \rightarrow \infty$. So instead the error band indicates the spread in results for different values of $L_t$.

\begin{figure}[htb]
\centering
\includegraphics[scale=1.0]{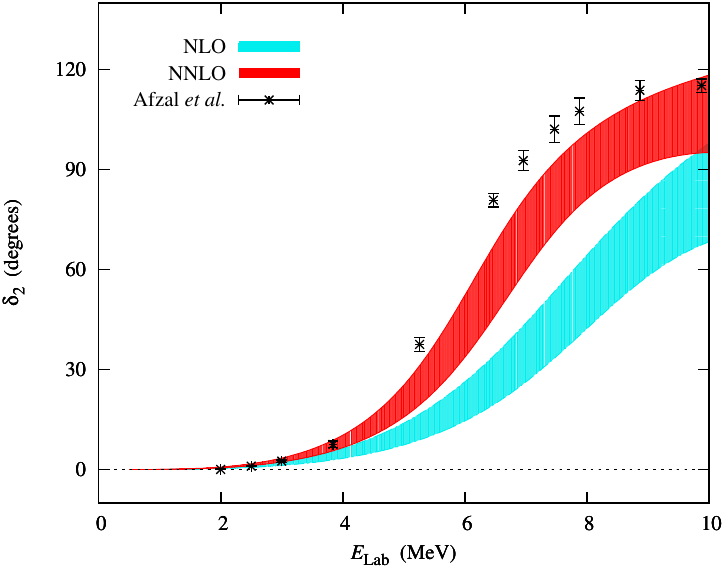}
\caption{$D$-wave phase shifts at NLO and NNLO and comparison with experimental
data \cite{Heydenburg:1956,Nilson:1958,Tombrello:1963,Afzal:1969}.}
\label{fig:D-wave}
\end{figure}

In Fig.~\ref{fig:D-wave} we show phase shifts for $D$-wave scattering
versus laboratory energy at NLO, NNLO, and the comparison with experimental data.    We find that the $2^+ $ resonance energy is pushed upwards by the Coulomb interactions and other higher-order corrections at NLO and NNLO.  The results at NNLO are in fairly good agreement with the experimental results. 

\begin{figure}[htb]
\centering
\includegraphics[scale=0.75]{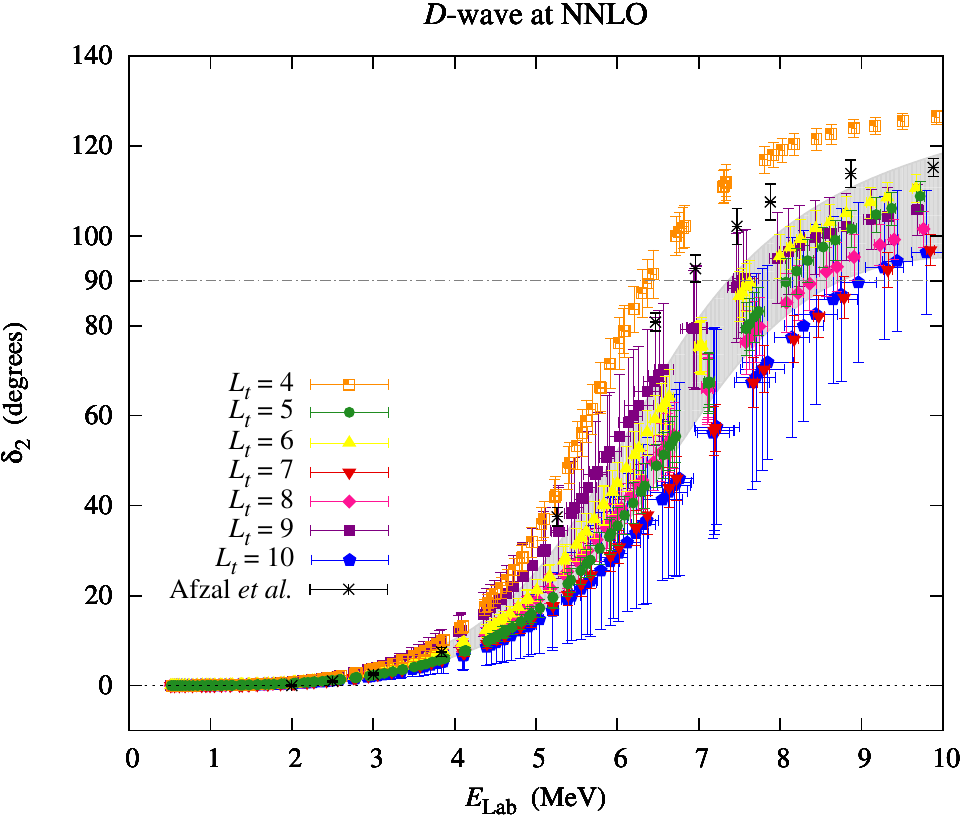}
\caption{$D$-wave phase shifts at NNLO for different numbers of projection
time steps $L_t$ and comparison with experimental
data \cite{Heydenburg:1956,Nilson:1958,Tombrello:1963,Afzal:1969}.}
\label{fig:D-wave_NNLO}
\end{figure}

In Fig.~\ref{fig:D-wave_NNLO} we plot the $D$-wave
phase shifts at NNLO for $L_t=4$ to $L_t =10$.  Again the error bars are computed using a jackknife analysis of the stochastic errors of the Monte Carlo data.  In this case we see a stronger systematic difference between even $L_t$ and odd
$L_t$ results, with the even $L_t$ data approaching from above and the odd $L_t$ data approaching from below.

\section{Summary and outlook}
In this letter we describe the first \textit{ab initio} calculation
of $^4\rm{He} + {^4\rm{He}}$ scattering.  We use lattice effective field theory and the adiabatic projection method to compute phase shifts for $S$-wave and $D$-wave scattering up to NNLO and find good agreement with experimental data. In order to perform these calculations, several new innovations were developed. This includes the use of spherical wave projections of the lattice initial states and a new algorithm that performs updates of both the auxiliary
field configurations and alpha cluster positions.
We plan to revisit these $^4\rm{He} + {^4\rm{He}}$ calculations again
in the future with different lattice spacings and going one order higher to NNNLO.  These phase shifts provide   useful benchmarks to assess systematic errors in calculations of higher-body nuclear systems.  
 

Perhaps the most significant outcome of this study is that there is now a new method for scattering and reactions that has very favorable scaling with particle number.  The computational
effort needed for the $A_1$-body + $A_2$-body problem is roughly $(A_1+A_2)^2$  for light and medium-mass nuclei, and there is no restriction on the projectile being very light. Since the problem of sign oscillations is greatly suppressed for alpha-like nuclei \cite{Chen:2004rq,Lee:2007eu,Lahde:2015ona}, this approach appears to be a viable  method to study important alpha processes involving heavier nuclei.

\section{Acknowledgements}
We acknowledge partial financial support from the 
Deutsche Forschungsgemeinschaft (Sino-German CRC 110), the Helmholtz Association
(Contract No.\ VH-VI-417), 
BMBF (Grant No.\ 05P12PDFTE), the U.S. Department of Energy (DE-FG02-03ER41260), and U.S. National Science Foundation grant No. PHY-1307453.
Further support
was provided by the EU HadronPhysics3 project and the ERC Project No.\ 259218
NUCLEAREFT. The computational resources 
were provided by the J\"{u}lich Supercomputing Centre at  Forschungszentrum
J\"{u}lich and by RWTH Aachen.

\bibliographystyle{apsrev}
\bibliography{References}
\end{document}